\documentclass[prd,superscriptaddress,onecolumn,showkeys]{revtex4}
\usepackage{eurosym}
\usepackage{amsfonts}
\usepackage{array}
\usepackage{amsthm}
\usepackage{bm}
\usepackage{palatino}
\usepackage{mathpazo}
\usepackage{amssymb}
\usepackage{eurosym}
\usepackage{amsmath}
\usepackage{epsfig}
\usepackage{graphics}
\usepackage{changes}
\usepackage{color}
\usepackage{graphicx}
\usepackage[colorlinks=true,
            linkcolor=blue,
           urlcolor=black,
           citecolor=black]{hyperref}

\def\be{\begin{equation}}
\def\ee{\end{equation}}
\def\beq{\begin{eqnarray}}
\def\eeq{\end{eqnarray}}

\def\bes{\begin{eqnarray}}
\def\ees{\end{eqnarray}}

\begin{document}

\title{\textbf{Tunneling Under the Influence of Quantum Gravity in Black Rings}}

\author{Riasat Ali}
\email{riasatyasin@gmail.com}
\affiliation{Department of Mathematics, GC
University Faisalabad Layyah Campus, Layyah-31200, Pakistan}

\author{Kazuharu Bamba}
\email{bamba@sss.fukushima-u.ac.jp}
\affiliation{Division of Human Support System, Faculty of Symbiotic Systems
Science, Fukushima University,
Fukushima 960-1296, Japan}

\author{Muhammad Asgher}
\email{m.asgher145@gmail.com}
\affiliation{Department of Mathematics, The Islamia
University of Bahawalpur, Bahawalpur-63100, Pakistan}

\author{Syed Asif Ali Shah}
\email{asifalishah695@gmail.com}
\affiliation{Department of Mathematics and Statistics, The University of Lahore 1-Km Raiwind Road,
Sultan Town Lahore 54000, Pakistan}

\begin{abstract}
We explore the Lagrangian equation in the background of generalized uncertainty principle.
The tunneling radiation through the black ring horizon is observed.
We investigated the tunneling radiation through the Hamilton-Jacobi method
for solutions of Einstein-Maxwell-dilation gravity theory.
The radiation of black ring without back reaction and self interaction of particles are studied.
Furthermore, we consider the quantum gravity effect on the stability of black ring.
\end{abstract}


\date{\today}
\maketitle

\section{Introduction}
The five dimensions rotating black ring solution has been analyzed in \cite{1}.
It is observed that the angular momentum and mass of black ring is same as a black hole.
Many researchers have analyzed the metric information for the well-known black rings \cite{2,3,4} and studied that the black ring with $S^{2}\times S^{1}$ horizon. 
The nature of the black ring space-time for the nonlinear $\sigma$-model and the normal horizon should be studied in \cite{5}.

The black ring evaporation as a result of tunneling radiation is one of the significant phenomenon.
The tunneling method for Dirac spin up/down particles for neutral black ring (NBR) five dimension spaces have been studied \cite{6}.
The investigation is established on tunneling to study the complex mathematical object for particle to go across with the black ring.
The NBR has horizon of the topology  $S^{1}\times S^{2}$.
The solution is describing a NBR, which takes necessity conic singularity at that region, there is no centrifugal force exist to develop equilibrium of NBR due to self gravity.
The tunneling radiation method by applying the Hamilton-Jacobi process and the WKB approximation of different types of particle of the black rings in five dimensions have been investigated \cite{7}.
It is suggested that to study the quantum method of tunneling radiation from the black rings.
Different phenomena are usually applied to study the particle action by computing its imaginary component.
The tunneling radiation of the particles and their the corresponding Hawking temperature of the black rings have been obtained.

The tunneling radiation from the different spin boson particle emission by applying the Hamilton Jacobi method to the Lagrangian field equation and evaluated the tunneling probability for the well-knows black holes \cite{8,9,10,11}.
Also, WKB approximation is used to a general black hole and computed the corresponding temperature.

The generalized uncertainty principle (GUP) plays a very significant purpose to analyze the gravitational effects.
To take the gravity effects of tunneling radiation and the Lagrangian equation will be modified by taking GUP effects.
The black hole and rotating black ring thermodynamics have furthermore been studied within the GUP effects \cite{12, 13, 14, 15, 16, 16b}.

The paper is sketched as: In Sec. \textbf{II}, we analyzed metric of the NBR and also studied the tunneling radiation from the NBR.
Section \textbf{III} provides tunneling radiation of dipole black ring (DBR) by using the same method in section \textbf{II}.
In Sec. \textbf{IV}, it contains the graphical behavior of temperature and analyze the GUP effect on temperature.
In Sec. \textbf{V} we have expressed the discussion and conclusions.

\section{Neutral Black Ring}
The quantum gravity caused by different spin particles which gets important at really high densities and keeps singularities in black rings.
The particle production is due to quantum gravity, every black ring may produce a new radiation on the inner and outer of its horizon.
Liu \cite{17} analyzed the black ring for first thermodynamic law and the emission probability is associated with the black ring entropy.

In this chapter, we discuss boson tunneling radiation for NBR five dimension spaces.
We study the Lagrangian equation with GUP by applying WKB approximation.
The study of black ring thermodynamics has implication in gravitational physics.
The metric of NBR  \cite{6} is given by

\begin{eqnarray}
ds^2&=&-\frac{G(y)}{G(x)}[dt-C(\lambda, \nu)R\frac{1+y}{G(y)}d\psi]^2+
\frac{R^2}{(y-x)^2}G(x)\nonumber\\
&&\left[-\frac{F(y)}{G(y)}d\psi^2-\frac{1}{F(y)}dy^2+\frac{1}{F(x)}dx^2+
\frac{F(x)}{G(x)}d\phi^2\right],\label{M}
\end{eqnarray}
and
\begin{eqnarray}
G(\xi)&=&\lambda\xi+1,~~~F(\xi)=(1+\nu\xi)(1-\xi^2),\nonumber\\
C(\lambda, \nu)&=&\sqrt{(\lambda-\nu)\lambda\frac{1+\lambda}{1-\lambda}}.\nonumber
\end{eqnarray}
The $\nu$ and $\lambda$ are dimensionless parameters and takes values in the range $(1>\lambda\geq\nu>0)$ and to expect the singularity of conical at $x=1$,
$\lambda$ and $\nu$ to be associated with each other, such that $\lambda=\frac{2\nu}{\nu^2+1}.$
The component $\psi$ and $\phi$ are two BR cycles, $x$ and $y$ takes the values $1\geq x\geq-1$ and $-1\geq y\geq-\infty$.
The horizon is around at $y=-\frac{1}{\nu}=y_{h}$. The $R$ has the dimension with fixed length.
The BR mass is  $M=\frac{3\pi \lambda R^2}{2(1-\nu)^2}$. 
In addition, there are three space-time coordinates $t$, $\phi$ and $\psi$ associated with Killing vectors and 5D space-time coordinates are taken as $x^{\mu}=(t, \phi, y, x, \psi)$. 
Next, we will analyze boson particles tunneling from the NBR.
The metric from Eq. (\ref{M}) can be rewritten as

\begin{equation}
ds^2=Udt^2+Vd\phi^2+Wdy^2+Xdx^2+Yd\psi^2+2Zdtd\psi,\label{M1}
\end{equation}
where
\begin{eqnarray}
U&=&-\frac{G(y)}{G(x)},~~~~
W=\frac{G(x)R^2}{(y-x)^{2}F(y)},~~~
V=\frac{F(x)R^2}{(x-y)^2},\nonumber\\
Y&=&-\frac{C^{2}(\lambda, \nu)(1+y)^{2}R^{2}}{G(x)G(y)}-
\frac{F(y)G(x)R^2}{(y-x)^{2}G(y)},\nonumber\\
X&=&\frac{G(x)R^2}{(y-x)^{2}F(x)},~~~Z=
\frac{C(\lambda, \nu)(1+y)R}{G(x)}.\nonumber
\end{eqnarray}
Now we focus on analyzing boson particles tunneling from the NBR. 
In curved space-time, boson particles should be satisfied with the following Lagrangian gravity equation without charge \cite{12,13} 

\begin{eqnarray}
&&\partial_{\mu}(\sqrt{-g}\chi^{\nu\mu})+
\sqrt{-g}\frac{m^2}{\hbar^2}\chi^{\nu}
+\beta \hbar^{2}\partial_{0}\partial_{0}
\partial_{0}(\sqrt{-g}g^{00}\chi^{0\nu})\nonumber\\
&-&\beta \hbar^{2}\partial_{i}\partial_{i}
\partial_{i}(\sqrt{-g}g^{ii}\chi^{i\nu})=0,\label{L}
\end{eqnarray}
and
\begin{eqnarray}
\chi_{\nu\mu}&=&(1-\beta{\hbar^2\partial_{\nu}^2})\partial_{\nu}\chi_{\mu}-
(1-\beta{\hbar^2\partial_{\mu}^2})\partial_{\mu}\chi_{\nu}.\nonumber
\end{eqnarray}
Here $\beta$, $m$ and $\chi$ are the quantum gravity parameter, particle mass and anti-symmetric tensor and $\partial_0$ and $\partial_i$ denotes the partial derivative with respect to $t$ and $i=(\phi, y, x, \psi)$, respectively.
The $\chi$ components are calculated as
\begin{eqnarray}
&&\chi^{0}=\frac{V}{UV-Z^2}\chi_{0}-\frac{Z}{UV-Z^2}\chi_{1},
~~\chi^{1}=\frac{-Z}{UV-Z^2}\chi_{0}-\frac{U}{UV-Z^2}\chi_{1},\nonumber\\
&&\chi^{2}=\frac{1}{W}\chi_{2},~~\chi^{3}=\frac{1}{X}\chi_{3},
~~\chi^{4}=\frac{1}{Y}\chi_{4},~~
\chi^{01}=\frac{Z^2\chi_{10}+UV\chi_{01}}{(UV-Z^2)^2},\nonumber\\
&&\chi^{02}=\frac{V\chi_{02}-Z\chi_{12}}{W(UV-Z^2)},~~
\chi^{03}=\frac{-Z\chi_{03}+U\chi_{13}}{X(UV-Z^2)},~~
\chi^{04}=\frac{-Z^2\chi_{04}+U\chi_{14}}{Y(UV-Z^2)},\nonumber\\
&&\chi^{12}=-\frac{Z\chi_{02}+U\chi_{12}}{W(UV-Z^2)},~~
\chi^{13}=\frac{-Z\chi_{03}+U\chi_{13}}{X(UV-Z^2)},~~
\chi^{14}=\frac{-Z\chi_{04}+U\chi_{14}}{Y(UV-Z^2)},\nonumber\\
&&\chi^{23}=\frac{1}{WX}\chi_{23},~~\chi^{24}=\frac{1}{WY}\chi_{24},
~~\chi^{34}=\frac{1}{XY}\chi_{34}.\nonumber
\end{eqnarray}
The WKB approximation is \cite{18}
\begin{equation}
\chi_{\nu}=c_{\nu}\exp[\frac{i}{\hbar}\Theta_{0}(t, \phi, y, x, \psi)+
\Sigma \hbar^{n}\Theta_{n}(t, \phi, y, x, \psi)].\nonumber
\end{equation}
Here,$(\Theta_{0},~~\Theta_{n})$ and $c_{\nu}$ are arbitrary functions and constant.
By neglecting the higher order terms for $n=1, 2, 3,...$ and applying Eq. (\ref{L}), we obtain the field equations and applying variables separation technique \cite{6}, we can take

\begin{equation}
\Theta_{0}=-Et+j\phi+I(x, y)+L\psi+K.
\end{equation}
Here, $E$ and $K$ are the particle energy and complex constant, $j$ and $L$ are denoting the angular momentum of particles and also associating to the directions $\phi$ and $\psi$ respectively.
From a set of field equations, we obtain a $5\times5$ equation of a matrix $G(c_{0},c_{1},c_{2}, c_{3}, c_{4})^T =0,$ the matrix elements should be expressed as

\begin{equation*}
G(c_{0},c_{1},c_{2},c_{3},c_{4})^{T}=0.
\end{equation*}
Which gives a $5\times5$ matrix presented as "$G$", whose elements are given as follows:

\begin{eqnarray}
G_{00}&=&Z^2\tilde{U}[J_{1}^{2}+\beta J_{1}^{4}]
-UV\tilde{U}[J_{1}^{2}+\beta J_{1}^{4}]-\frac{V}{W}[I_{1}^{2}+\beta I_{1}^{4}]
+\frac{Z}{X}[I_{2}^{2}+\beta I_{2}^{4}]\nonumber\\
&+&\frac{Z}{Y}[L_{1}^{2}+\beta L_{1}^{4}]-m^2 V,\nonumber\\
G_{01}&=&Z^2\tilde{U}[J_{1}+\beta J_{1}^{3}]E
-UV\tilde{U}[J_{1}+\beta J_{1}^{3}]E+\frac{Z}{W}[I_{1}^{2}+\beta I_{1}^{4}]+\frac{U}{X}[I_{2}^{2}+\beta I_{2}^{4}]\nonumber\\
&-&\frac{U}{Y}[L_{1}^{2}+\beta L_{1}^{4}]-m^2 Z,\nonumber\\
G_{02}&=&-\frac{V}{W}[E+\beta E^3]I_{1}-\frac{Z\tilde{U}}{W}
[J_{1}+\beta J_{1}^{3}]I_{1},\nonumber\\
G_{03}&=&\frac{V}{X}[E+\beta E^3]I_{2}+\frac{U\tilde{U}}{X}
[J_{1}+\beta J_{1}^{3}]I_{2},\nonumber\\
G_{04}&=&\frac{Z}{Y}[E+\beta E^3]L_{1}+\frac{U\tilde{U}}{Y}
[J_{1}+\beta J_{1}^{3}]L_{1},\nonumber\\
G_{10}&=&Z^2\tilde{U}[J_{1}+\beta J_{1}^{3}]E-
UV\tilde{U}[J_{1}+\beta J_{1}^{3}]E+\frac{Z}{W}[I_{1}^{2}+\beta I_{1}^{4}]+\frac{Z}{X}
[I_{2}^{2}+\beta I_{2}^{4}]\nonumber\\
&+&\frac{Z}{Y}[L_{1}^{2}+
\beta L_{1}^{4}]+m^2 Z,\nonumber
\end{eqnarray}
\begin{eqnarray}
G_{11}&=&Z^2\tilde{U}[E^2+\beta E^4]
-UV\tilde{U}[E^2+\beta E^4]-\frac{U}{W}[I_{1}^{2}+\beta I_{1}^{4}]-\frac{U}{X}[I_{2}^{2}+\beta I_{2}^{4}]I_{2}\nonumber\\
&-&\frac{U}{Y}[L_{1}^{2}+\beta L_{1}^{4}]+m^2 U,\nonumber\\
G_{12}&=&\frac{Z}{W}[E+\beta E^3]I_{1}
+\frac{U\tilde{U}}{W}[J_{1}+\beta J_{1}^{3}]I_{1},\nonumber\\
G_{13}&=&\frac{Z}{X}[E+\beta E^3]I_{1}+\frac{U\tilde{U}}{X}
[J_{1}+\beta J_{1}^{3}]I_{2},\nonumber\\
G_{14}&=&\frac{Z}{Y}[E+\beta E^3]L_{1}+\frac{U\tilde{U}}{Y}
[J_{1}+\beta J_{1}^{3}]L_{1},\nonumber\\
G_{20}&=&-V\tilde{U}[I_{1}+\beta I_{1}^{3}]E
-Z\tilde{U}[I_{1}+\beta I_{1}^{3}]J_{1},\nonumber\\
G_{21}&=&Z\tilde{U}[I_{1}+\beta I_{1}^{3}]E+U\tilde{U}
[I_{1}+\beta I_{1}^{3}]J_{1},\nonumber\\
G_{22}&=&-V\tilde{U^2}[E^2+\beta E^4]-Z\tilde{U^2}[J_{1}+\beta J_{1}^{3}]E-U\tilde{U}[J_{1}^{2}+\beta J_{1}^{4}]-m^2\nonumber\\
&-&\frac{1}{X}[I_{2}^{2}+\beta I_{2}^{4}]-\frac{1}{Y}
[L_{1}^{2}+\beta L_{1}^{4}],\nonumber\\
G_{23}&=&\frac{1}{X}[I_{1}+\beta I_{1}^{3}]I_{2},~~
G_{24}=\frac{1}{Y}[I_{1}+\beta I_{1}^{3}]L_{1},\nonumber\\
G_{30}&=&-V\tilde{U}[I_{2}+\beta I_{2}^{3}]E
-Z\tilde{U}[I_{2}+\beta I_{2}^{3}]J_{1},\nonumber\\
G_{31}&=&Z\tilde{U}[I_{2}+\beta I_{2}^{3}]E
+U\tilde{U}[I_{2}+\beta I_{2}^{3}]J_{1},\nonumber\\
G_{32}&=&\frac{1}{W}[I_{2}+\beta I_{2}^{3}]I_{1},\nonumber\\
G_{33}&=&-V\tilde{U}[E^2+\beta E^4]-Z\tilde{U}
[J_{2}+\beta J_{1}^{4}]-\frac{1}{W}[I_{1}^{2}+\beta I_{1}^{4}]-\frac{1}{Y}
[L_{1}^{2}+\beta L_{1}^{4}]-m^2,\nonumber\\
G_{34}&=&\frac{1}{Y}[I_{2}+\beta I_{2}^{3}]L_{1},\nonumber\\
G_{40}&=&-V\tilde{U}[L_{1}+\beta L_{1}^{3}]E-Z\tilde{U}
[L_{1}+\beta L_{1}^{3}]J_{1},\nonumber\\
G_{41}&=&Z\tilde{U}[L_{1}+\beta L_{1}^{3}]E
+U\tilde{U}[L_{1}+\beta L_{1}^{3}]J_{1},\nonumber\\
G_{42}&=&\frac{1}{W}[L_{1}+\beta L_{1}^{3}]I_{1},~~G_{43}=\frac{1}{X}
[L_{1}+\beta L_{1}^{3}]I_{2},\nonumber\\
G_{44}&=&-V\tilde{U}[E^2+\beta E^4]-Z\tilde{U}[J_{1}+\beta J_{1}^{3}]
-Z\tilde{U}[E+\beta E^3]J_{1}-U\tilde{U}[J_{1}^{2}+\beta J_{1}^{4}]
\nonumber\\
&&-\frac{1}{W}[I_{1}^{2}+\beta I_{1}^{4}]-\frac{1}{X}
[I_{2}^{2}+\beta I_{2}^{4}]-m^2,\nonumber
\end{eqnarray}
where $\tilde{U}=\frac{1}{UV-Z^2},~J_{1}=\partial_{\phi}\Theta_{0}$, $I_{1}=\partial_{x}\Theta_{0}$,
$I_{2}=\partial_{y}\Theta_{0}$
and $L_{1}=\partial_{\psi}\Theta_{0}$.
For the non-trivial result, the determinant $\textbf{G}$ is
zero and we get\\
\begin{equation}\label{I}
ImI_{\pm}=\pm \int\sqrt{\frac{E^{2}+X_{1}
[1+\beta\frac{X_{2}}{X_{1}}]}{-\frac{UV-Z^2}{VX}}}dy,
\end{equation}
where, positive and negative sign denotes the radical functions of outgoing
and ingoing boson particles respectively, while $X_{1}$ and $X_{2}$ functions can be defined as

\begin{eqnarray}
X_{1}&=&-ZXE\tilde{U}J_{1}-U\tilde{U}X J_{1}^{2}-
\frac{X}{Y}L_{1}^{2}-m^2X\nonumber\\
X_{2}&=&-VXE^4\tilde{U}-ZXE\tilde{U}J_{1}^{3}-
U\tilde{U}X J_{1}^4-I_{2}^{4}-\frac{X}{Y}L_{1}^{4}\nonumber
\end{eqnarray}
Extending the A(y) and B(y) functions in Taylor's series near the NBR horizon, we get

\begin{equation}
A(y_{+})\simeq \acute{A}(y_{+})(y-y_{+}),~~~~B(y_{+})\simeq
\acute{B}(y_{+})(y-y_{+}).\label{AB}
\end{equation}
Applying the Eq. (\ref{AB}) in Eq. (\ref{I}), one can take that the resulting solution has pole at $y=y_{+}$.
For the computation of the temperature by applying tunneling phenomenon, it is assumed that
to take the singularity by particular contour around the pole.
In our investigation, the coordinates of the NBR metric, the tunnel of outgoing
boson particles can be found by assuming an infinitesimal semi-circle under the
pole $y=y_{+}$, as the incoming boson particles this contour is assumed over the pole.
Since applying Eqs. (\ref{I}) and (\ref{AB}), integrating the lead field equation around the NBR pole, we have

\begin{equation}
ImI_{\pm}=\pm\frac{i\pi E}{2\kappa(y_{+})}[1+\Xi \beta]\label{im}.
\end{equation}

The surface gravity of NBR is given as
\begin{equation}
\kappa(y_{+})=\frac{R(\lambda x+1)\sqrt{\lambda[a^2(-2y+\nu-3y^2)+2ab]}}
{2(\lambda y+1)a^2b}.\label{k}
\end{equation}
Here, $a=x-y$ and $b=y^2-\nu y+\nu y^3-1$,
as boson particles taking spin-1, when $y$ direction measuring spin
and there would be two kinds, one is spin up kind and which exist the
like $y$ direction, the other spin down kind take the opposite direction\\
$(I_{+}(y)=-I_{-}(y))$.

The tunneling probability of NBR is given as
\begin{eqnarray}
\Gamma &=&\frac{\Gamma_{emission}}{\Gamma_{absorption}}=
\frac{\exp(-2ImI_{+}(x)-2ImI_{+}(y)-2Im\Theta)}
{\exp(-2ImI_{-}(x)-2ImI_{-}(y)-2Im\Theta)}=\exp{\left(-4ImI_{+}(y)\right)}\nonumber\\
&=&\exp\left(-\frac{4\pi(\lambda y+1)a^2 b E[1+\Xi \beta]}
{R(\lambda x+1)\sqrt{\lambda[a^2(-2y+\nu-3y^2)+2ab]}}\right).\label{p}
\end{eqnarray}
It should be calculated that Eq. (\ref{im}) gives near the horizon of the NBR along $y$ direction.

By equating Eq.(\ref{p}) formulation with the factor of Boltzmann $\exp[-\beta_{1} E]$, one can deduce the temperature which is
$T_{H}=\frac{1}{\beta_{1}}$ at the outer NBR horizon $y$.
For this result, we can obtain the NBR temperature as

\begin{equation}
T_{H}=\frac{R(\lambda x+1)\sqrt{\lambda[a^2(-2y+\nu-3y^2)+2ab]}}
{\pi(\lambda y+1)a^2b}[1+\beta \Xi]^{-1}.\label{T}
\end{equation}
The Hawking of NBR depends upon these parameters $\nu$, $R$, $\beta$ and $\lambda$.
The resulting temperature at which boson particle tunnel by the horizon is different to the temperature of a charge boson particle at which they tunnel through the NBR horizon.
It is observed that the resulting Hawking temperature (\ref{T}) is just for spin up boson particles.
For spin up case, assuming a way fully corresponds to the spin down case solution, but in the opposite direction which means both spin down and spin up boson particles are radiated at the like rate.

\section{Dipole Black Ring}

In this section, we will analyze Hawking temperature of boson
particles through the tunneling process from the DBR. The DBR
contribution the like action as NBR, since they physically obtain
more similar behavior. We have analyzed that, there is an important
physical object DBR from the gravity theory. The five dimensions DBR
was constructed in \cite{6} and its metric assumes in the form as

\begin{eqnarray}
ds^2&=&-\frac{G(y)}{G(x)}\left(\frac{K(x)}{K(y)}\right)^{\frac{N}{3}}
\left[dt-C(\lambda, \nu)R\frac{y+1}{G(y)} d\psi\right]^2\nonumber\\
&&+\frac{R^2}{(x-y)^2}G(x)\left(K(x)K^2(y)\right)^{\frac{N}{3}}\nonumber\\
&&\times\left[\frac{F(y)}{G(y)
K^N(y)}d\psi^2-\frac{dy^2}{F(y)}+\frac{dx^2}{F(x)}
+\frac{F(x)}{G(x)K^N(x)}d\phi^2\right],\label{D1}
\end{eqnarray}
where, $C(\lambda, \nu)$, $F(\xi)$ and $G(\xi)$ are the similar for NBR and $K(\xi)=1-\xi\mu$\\ $(1>\mu\geq 0).$
The constant of dilation coupling is associated with the dimensionless constant as $\alpha^2=(\frac{4}{N}-\frac{4}{3})(3\geq N>0).$
The event horizon of DBR is located at $y=-\frac{1}{\nu}=yK$.
We analyze the tunneling and temperature for boson particles in the horizon of DBR.

The tunneling rate of boson particle in DBR horizon is given as
\begin{eqnarray}
\acute{\Gamma} &=&\exp\left(-4\pi\frac{ E[1+\Xi \beta]}{\sqrt{fd(g+h)(e+4ab-2al)}}\right).\label{P}
\end{eqnarray}

Here, $a=x-y$,~~ $b=y^2-\nu y+\nu y^3-1$,~~$d=R^2(\lambda x+1)[(1-\nu x)(1-\nu y)]^{\frac{N}{3}}$,
~~$e=\frac{-2N \nu}{3(1-\nu y)}$,~~$f=(\frac{1-\nu x}{1-\nu y})^{\frac{N}{3}},$
~~$g=\frac{N\nu(\lambda y+1)}{3(\lambda x+1)(1-\nu y)}$ and $h=\frac{\lambda}{\lambda x+1}.$\\\\\\

Now we compute the Hawking temperature,
\begin{eqnarray}
\acute{T}_{H}&=&\frac{\sqrt{fd(g+h)(e+4ab-2al)}}
{\pi}[1+\Xi \beta]^{-1}.\label{t}
\end{eqnarray}
This solution has been computed by above applies Hamilton-Jacobi phenomenon and boson particles out the horizon of DBR are only for vector cases.
We are only assuming the case of boson particles with spin up. In our investigation, assuming a same method and we will compute the like solution for boson particles with spin down but opposite direction.

\section{Gravitational Effect on Temperature}
In this section, we study the physical importance of quantum gravity and also analyze the impact of quantum gravity parameter on the instability and stability of NBR and DBR.

\subsection{$T_{H}$ versus $y$}

In this subsection, we can take quantum gravity value in the range of $100 \leq \beta\leq 300$.

\begin{itemize}
\item{\bf{Figure~1}} indicates the graph between T and y for varying quantum gravity and fixed values for other parameters.\\
(i). We analyzed that the gravity parameter decreases for a small variance of Hawking temperature.\\
(ii). We examined that the quantum gravity gradually decreasing and then finally goes to a large change of $\beta$ which is the unstable condition of NBR.\\
So, we observed that for small values of $\beta$ the NBR is stable but as $\beta$ increases it indicates the unstable behavior of NBR.
\item{\bf{Figure~2}}, shows the behavior of temperature by varying gravity parameter.\\
(i). We observed that the Hawking temperature decreasing constantly for small range values of $\beta$.\\
(ii). As gravity increases the Hawking temperature gradually increases and then approximation goes to infinity.
So, it is observed that in the range of $0\leq y\leq 2.7$, the DBR is stable and
Hawking temperature remains constant and the range of $2.7< y< \infty$ the DBR is unstable.
\end{itemize}
We observed that the graphical behavior of decrease and increase Hawking temperature when $\beta<100$ and  $\beta>100$, respectively. As, for $\beta = 100$ the temperature remains constant in a particular range of $y$.\\\\\\\\\\\\
\begin{figure}\begin{center}
\epsfig{file=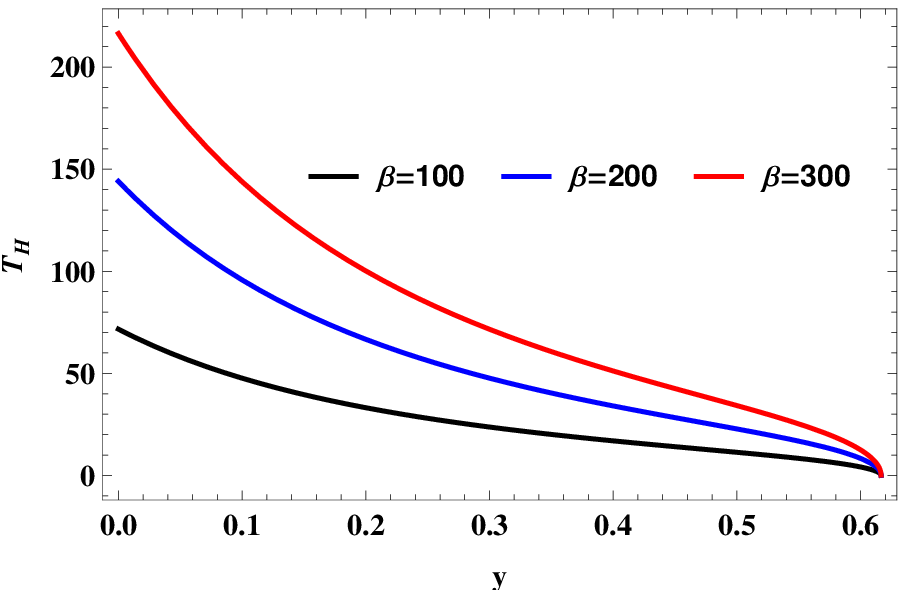, width=0.75\linewidth}
\caption{$T_{H}$ versus $y$ for $R = 1,~~\lambda = 0.9,
~~\nu = 0.75,~~x = -0.5$ and $\Xi=1.$}
\end{center}
\begin{center}
\epsfig{file=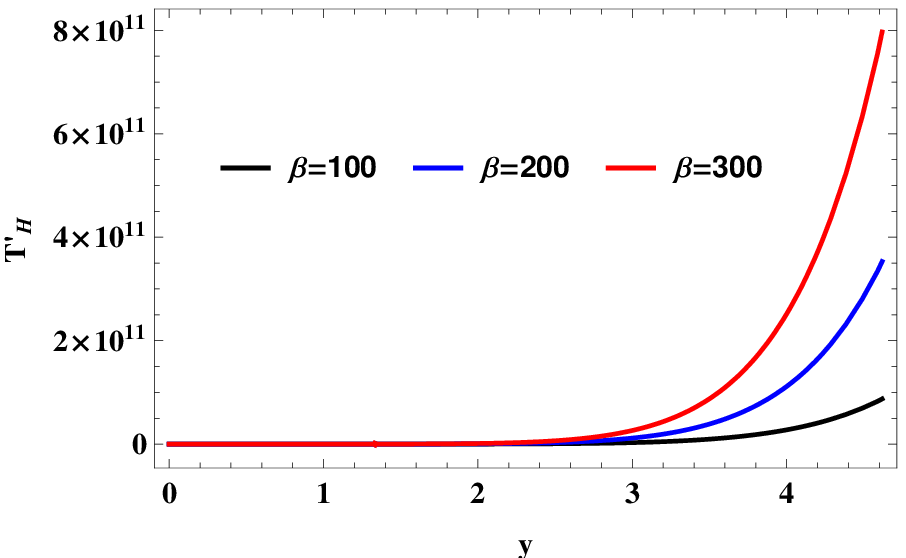, width=0.82\linewidth}
\caption{$T^{\prime}_{H}$ versus $y$ for $R = 1,~~N=3,
~~\lambda = 0.9,~~\nu = 0.75,~~x = -0.5$ and $\Xi=1.$}
\end{center}
\end{figure}

\section{Conclusions}

The tunneling spectrum of the Dirac and boson particles
for black ring space-time have been evaluated in different coordinate
system \cite{6,7}. The coordinate system in the rotating frame does not affect the value of $T_H$.
We evaluated boson particulate tunneling
from the black rings to the Lagrangian gravity equation
using the WKB approximation to the Hamilton-Jacobi phenomenon.

The WKB approximation tells us the tunneling probability for the classical forbidden trajectory from inner to outer horizon is associated with imaginary part of the emitted boson particles action across the black rings horizon.
Firstly, we analyze the imaginary part of the action, as we know that the
tunneling rate is proportional to the imaginary part of the particle action.
Moreover, we examine the comparison of Hawking temperature values by assuming Boltzmann
factor in both cases and analysis of the spectra in general.
We consider both the cases of boson particles with spin-up and spin-down in the $y$-direction.
Tunneling of radiation by assuming conservation of energy-charge
and quantum gravity is studied.

For 5D black ring, the $T_H$ given by equation (\ref{T}) related to the $\lambda$,
$x$, $y$, $a$, $b$, $R$, $\nu$ and $\beta$, which is similar to
the $\acute{T}_{H}$ as given in Eq. (\ref{t}) when $N=3$ and $\mu=0$.
The modified temperatures are associated with the quantum gravity and geometry of the black ring.
In the absence of quantum gravity, the normal tunneling and
normal temperature are obtained. The Hawking temperature rises with the lowering of the horizon in Fig.
(1) (which is physical) has been analyzed. Moreover, we have studied the
Hawking temperature increases as the horizon increases in Fig. (2) (which is non physical).

\section*{Acknowledgments}
The work of KB was supported in part by the JSPS KAKENHI Grant Number JP
25800136 and Competitive Research Funds for Fukushima University Faculty
(19RI017).

\end{document}